\documentclass[prd, preprint, nofootinbib]{revtex4}

\usepackage{graphicx}
\usepackage{amsfonts}
\usepackage{latexsym}
\usepackage{amsmath}
\usepackage{amssymb}
\usepackage{epsfig}

\begin{document}

\title{ Probing the seesaw scale with gravitational waves}

\author{Nobuchika Okada}
 \email{okadan@ua.edu}
 \affiliation{
Department of Physics and Astronomy, 
University of Alabama, Tuscaloosa, Alabama 35487, USA}

\author{Osamu Seto}
 \email{seto@particle.sci.hokudai.ac.jp}
 \affiliation{Institute for International Collaboration, Hokkaido University, Sapporo 060-0815, Japan}
 \affiliation{Department of Physics, Hokkaido University, Sapporo 060-0810, Japan}

%\date{\today}
%

%%%%%%%%%%%%%%%%%%%%%%
\begin{abstract}
%%%%%%%%%%%%%%%%%%%%%%
The $U(1)_{B-L}$ gauge symmetry is a promising extension of the standard model of particle physics, 
  which is supposed to be broken at some high energy scale. 
Associated with the $U(1)_{B-L}$ gauge symmetry breaking, right-handed neutrinos acquire their Majorana masses 
  and then tiny light neutrino masses are generated through the seesaw mechanism. 
In this paper, we demonstrate that the first-order phase transition of the $U(1)_{B-L}$ gauge symmetry breaking 
  can generate a large amplitude of stochastic gravitational wave (GW) radiation for some parameter space of the model,  
  which is detectable in future experiments. 
Therefore, the detection of GWs is an interesting strategy to probe the seesaw scale
  which can be much higher than the energy scale of collider experiments. 
%%%%%%%%%%%%%%%%%%%%%%
\end{abstract}
%%%%%%%%%%%%%%%%%%%%%%

%\pacs{}

\preprint{EPHOU-18-007} 

\vspace*{3cm}
\maketitle

%==================================%
%          Main body               %
%==================================%

%%%%%%%%%%%%%%%%%%%%%%%
\section{Introduction}
%%%%%%%%%%%%%%%%%%%%%%%

The nonvanishing neutrino masses have been established through various neutrino oscillation phenomena. 
The most attractive idea to explain the tiny neutrino masses is the so-called seesaw mechanism  
 with heavy Majorana right-handed (RH) neutrinos~\cite{Seesaw}. 
Then, the origin of neutrino masses is ultimately reduced to questions
 on the origin of RH neutrino masses.
It is natural to suppose that masses of RH neutrinos are also generated associated with developing
 the vacuum expectation value (VEV) of a Higgs field which breaks a certain (gauge) symmetry at a high energy scale.

As a promising and minimal extension of the standard model (SM), we may consider models 
 based on the gauge group $SU(3)_C \times SU(2)_L \times U(1)_Y \times U(1)_{B-L}$~\cite{Mohapatra:1980qe} 
 where the $U(1)_{B-L}$ (baryon number minus lepton number) gauge symmetry is supposed to be broken 
 at a high energy scale.  
In this class of models with a natural/conventional $U(1)_{B-L}$ charge assignment,
 the gauge and gravitational anomaly cancellations require us to introduce three RH neutrinos
 whose Majorana masses are generated by the spontaneous breakdown of the $U(1)_{B-L}$ gauge symmetry. 
In the case that the $U(1)_{B-L}$ symmetry breaking takes place at an energy scale higher than the TeV scale,
 it is very difficult for any collider experiments to address the mechanism of the symmetry breaking and
 the RH neutrino mass generation.

Detection of gravitational waves brings information about the evolution of the very early Universe. 
Cosmological GWs could originate from, for instance, quantum fluctuations during inflationary
 expansion~\cite{Starobinsky:1979ty} and phase transitions~\cite{Witten:1984rs,Hogan:1986qda}.  
If a first-order phase transition occurs in the early Universe,
 the dynamics of bubble collision~\cite{Turner:1990rc,Kosowsky:1991ua,Kosowsky:1992rz,Turner:1992tz,Kosowsky:1992vn}
 and subsequent turbulence of the plasma~\cite{Kamionkowski:1993fg,Kosowsky:2001xp,Dolgov:2002ra,Gogoberidze:2007an,Caprini:2009yp}
 and sonic waves generate GWs~\cite{Hindmarsh:2013xza,Hindmarsh:2015qta,Hindmarsh:2016lnk}.  
These might be within a sensitivity of future space interferometer experiments such as eLISA~\cite{Seoane:2013qna}; the Big Bang Observer (BBO)~\cite{Harry:2006fi} and DECi-hertz Interferometer Observatory (DECIGO)~\cite{Seto:2001qf}; or even ground-based detectors such as
Advanced LIGO (aLIGO)~\cite{Harry:2010zz}, KAGRA~\cite{Somiya:2011np} and VIRGO~\cite{TheVirgo:2014hva}. 

The spectrum of stochastic GWs produced by the first-order phase transition in the early Universe,
 in particular, by the SM Higgs doublet field, has been investigated in the literature. 
Here, the phase transition occurs at the weak scale.    
See, for instance, Ref.~\cite{Caprini:2018mtu} for a recent review.

In this paper, we focus on GWs from the first-order phase transition associated 
 with the spontaneous $U(1)_{B-L}$ gauge symmetry breaking at a scale higher than the TeV scale.
GWs generated by a $U(1)_{B-L}$ extended model with the classical conformal invariance~\cite{Iso:2009ss,Iso:2009nw},
 where its phase transition takes place around the weak scale, have been studied in Ref.~\cite{Jinno:2016knw}.
GWs from a second-order $B-L$ phase transition during reheating have been studied in Ref.~\cite{Buchmuller:2013lra}.
In this paper, we consider a slightly extended Higgs sector from the minimal model 
  and introduce an additional $U(1)_{B-L}$ charged Higgs field with its charge $+1$.  
This is one of the key ingredients in this paper. 
GWs generated by a phase transition in this extended scalar potential, but at TeV scale, 
 have been studied in Ref.~\cite{Chao:2017ilw}.
As we will show below, the new Higgs field plays a crucial role in causing
 the first-order phase transition 
 of the $U(1)_{B-L}$ symmetry breaking and the amplitude of resultant GWs generated 
 by the phase transition can be much larger than the one we naively expect.

%%%%%%%%%%%%%%%%%%%%%%%%%%%%%%%%%%%%%%%%%%%%%%%%%%%%%%
\section{GW generation by a cosmological first-order phase transition}
%%%%%%%%%%%%%%%%%%%%%%%%%%%%%%%%%%%%%%%%%%%%%%%%%%%%%%

In this section, we briefly summarize the properties of GWs produced by a first-order phase transition
  in the early Universe. 
There are three main GW production processes and mechanisms:
 bubble collisions, turbulence~\cite{Kamionkowski:1993fg} and
 sound waves after bubble collisions~\cite{Hindmarsh:2013xza}.
The GW spectrum generated by a first-order phase transition is mainly characterized by two quantities:
 the ratio of the latent heat energy to the radiation energy density, which is expressed by
 a parameter $\alpha$ and the transition speed $\beta$ defined below.
In this section, we introduce those parameters and the fitting formula
 of the GW spectrum.

%%%%%%%%%%%%%%%%%%%%%%%%%%%%%%%%%%
\subsection{Scalar potential parameters related to the GW spectrum}
%%%%%%%%%%%%%%%%%%%%%%%%%%%%%%%%%%

We consider the system composed of radiation and a scalar field $\phi$ at temperature $T$.
The energy density of radiation is given by
\begin{equation}
\rho_\mathrm{rad} = \frac{\pi^2 g_*}{30}T^4,
\end{equation}
 with $g_*$ being the number of relativistic degrees of freedom in the thermal plasma.
At the moment of a first-order phase transition, the potential energy of the scalar field includes
 the latent energy density given by
\begin{equation}
\epsilon = \left.\left(V - T\frac{\partial V}{\partial T}\right)\right|_{\{\phi_\mathrm{high}, T_{\star}\}} - \left.\left(V -T\frac{\partial V}{\partial T}\right)\right|_{\{\phi_\mathrm{low}, T_{\star}\}}, 
\end{equation}
 where $\phi_{\mathrm{high}(\mathrm{low})}$ denotes the field value of $\phi$ at the high (low) vacuum.  
Here and hereafter, quantities with the subscript $\star$ stand for those
 at the time when the phase transition takes place~\cite{Huber:2008hg}.
Then, a parameter $\alpha$ is defined as
\begin{equation}
\alpha \equiv  \frac{\epsilon}{\rho_\mathrm{rad}} .  
\end{equation}

The bubble nucleation rate per unit volume at a finite temperature is given by
\begin{equation}
\Gamma(T) =  \Gamma_0 e^{-S(T)} \simeq \Gamma_0 e^{-S^3_E(T)/T} ,  
\end{equation}
 where $\Gamma_0$ is a coefficient of the order of the transition energy scale,
 $S$ is the action in the four-dimensional Minkowski space, 
 and $S^3_E$ is the three-dimensional Euclidean action~\cite{Turner:1992tz}.
The inverse of the transition timescale can be defined as
\begin{equation}
\beta \equiv -\left.\frac{d S}{d t}\right|_{t_\star}. 
\end{equation}
Its dimensionless parameter $\beta/H_{\star}$ can be expressed as
\begin{align}
\frac{\beta}{H_{\star}} \simeq \left. T\frac{d S}{d T}\right|_{T_{\star}} = \left. T\frac{d (S^3_E/T)}{d T} \right|_{T_{\star}} . 
\end{align}

\subsection{GW spectrum}

\subsubsection{Bubble collisions}

Under the envelope approximation\footnote{For a recent development beyond the envelope approximation, see Ref.~\cite{Jinno:2016vai}.} 
 and for $\beta/H_{\star} \gg 1$~\cite{Kosowsky:1992vn}, 
 the peak frequency and the peak amplitude of GWs generated by bubble collisions are given by
\begin{align}
f_\mathrm{peak} &\simeq 17 \left(\frac{f_{\star}}{\beta}\right) \left(\frac{\beta}{H_{\star}}\right)
 \left(\frac{T_{\star}}{10^8 \, \mathrm{GeV}}\right)\left(\frac{g_*}{100}\right)^{1/6} \mathrm{Hz} , \\
h^2 \Omega_{GW}(f_\mathrm{peak} ) &\simeq 1.7 \times 10^{-5} \kappa^2\Delta \left(\frac{\beta}{H_{\star}}\right)^{-2}
 \left(\frac{\alpha}{1+\alpha}\right)^2 \left(\frac{g_*}{100}\right)^{-1/3} ,
\end{align}
with the following fitting functions
\begin{align}
\Delta &= \frac{0.11 v_b^3}{0.42+v_b^2},\\
\frac{f_{\star}}{\beta} &= \frac{0.62}{1.8-0.1 v_b+v_b^2},
\end{align}
 where $v_b$ denotes the bubble wall velocity.
The efficiency factor ($\kappa$) is given by~\cite{Kamionkowski:1993fg}
\begin{equation}
\kappa = \frac{1}{1+ A \alpha}\left( A \alpha +\frac{4}{27}\sqrt{\frac{3\alpha}{2}} \right),
\end{equation}
 with $A = 0.715 $.
The full GW spectrum is expressed as~\cite{Huber:2008hg}
\begin{equation}
\Omega_{GW}(f)  = \Omega_{GW}(f_\mathrm{peak})
 \frac{(a+b)f_\mathrm{peak}^b f^a}{ b f_\mathrm{peak}^{a+b} + a f^{a+b}},
\end{equation}
 with numerical factors $a \in [2.66, 2.82]$ and $b \in [0.90, 1.19]$.
We set the values of $(a, b, v_b)=(2.7, 1.0, 0.6)$ in our analysis.

%%%%%%%%%%%%%%%%%
\subsubsection{Sound waves}
%%%%%%%%%%%%%%%%%
 
The peak frequency and the peak amplitude of GWs generated by sound waves are given
 by~\cite{Hindmarsh:2013xza,Hindmarsh:2015qta} 
\begin{align}
f_\mathrm{peak} &\simeq 19 \frac{1}{v_b} \left(\frac{\beta}{H_{\star}}\right)
 \left(\frac{T_{\star}}{10^8 \, \mathrm{GeV}}\right)\left(\frac{g_*}{100}\right)^{1/6} \mathrm{Hz} , \\
h^2\Omega_{GW}(f_\mathrm{peak} ) &\simeq 2.7 \times 10^{-6} \kappa_v^2 v_b \left(\frac{\beta}{H_{\star}}\right)^{-1}
 \left(\frac{\alpha}{1+\alpha}\right)^2 \left(\frac{g_*}{100}\right)^{-1/3} .
\end{align}
The efficiency factor ($\kappa_v$) is given by~\cite{Espinosa:2010hh}
\begin{equation}
\kappa_v \simeq \left\{
\begin{array}{lll}
  v_b^{6/5} \frac{6.9 \alpha}{1.36-0.037\sqrt{\alpha}+\alpha }  \quad & \textrm{for} & v_b \ll c_s \\
%  \frac{\alpha^{2/5}}{0.017+( 0.997+{\alpha})^{2/5} }    & \textrm{for} & v_b \simeq c_s \\
%  \frac{ \sqrt{\alpha} }{0.135+ \sqrt{0.98 + \alpha} }   & \textrm{for} & v_b \simeq v_J \\
  \frac{\alpha}{0.73 + 0.083\sqrt{\alpha}+\alpha }       & \textrm{for} & v_b \simeq 1 \\
\end{array}
\right. ,
\end{equation}
 with $c_s$ being the sonic speed.
The spectrum shape is expressed as~\cite{Caprini:2015zlo}
\begin{equation}
\left(\frac{f}{f_{\mathrm{peak}}} \right)^3 
 \left( \frac{7}{ 4+ 3 \left(\frac{f}{f_\mathrm{peak}}\right)^2 } \right)^{7/2} .
\end{equation}

\subsubsection{Turbulence}

The peak frequency and amplitude of GWs generated by turbulence are given by~\cite{Kamionkowski:1993fg}
\begin{align}
f_\mathrm{peak} &\simeq 27 \frac{1}{v_b} \left(\frac{\beta}{H_{\star}}\right)
 \left(\frac{T_{\star}}{10^8 \, \mathrm{GeV}}\right)\left(\frac{g_*}{100}\right)^{1/6} \mathrm{Hz} , \\
h^2\Omega_{GW}(f_\mathrm{peak} ) &\simeq  3.4 \times 10^{-4} v_b \left(\frac{\beta}{H_{\star}}\right)^{-1}
 \left(\frac{\kappa_\mathrm{turb} \alpha}{1+\alpha}\right)^{3/2}
 \left(\frac{g_*}{100}\right)^{-1/3} .
\end{align}
In our analysis, we conservatively set the efficiency factor for turbulence to be 
 $\kappa_\mathrm{turb} \simeq 0.05 \kappa_v$ as in Ref.~\cite{Caprini:2015zlo}.
The spectrum shape is given by~\cite{Caprini:2009yp,Binetruy:2012ze,Caprini:2015zlo}
\begin{equation}
\frac{\left(\frac{f}{f_{\mathrm{peak}}} \right)^3}{
 (1 + \frac{f}{f_\mathrm{peak}} )^{11/3}(1+\frac{8\pi f}{h_{\star}})  }  ,
\end{equation}
with
\begin{equation}
h_{\star} = 17 \left(\frac{T_{\star}}{10^8 \mathrm{GeV}}\right)\left(\frac{g_*}{100}\right)^{1/6} \mathrm{Hz} .
\end{equation}
%

%%%%%%%%%%%%%%%%%%%%%%%
\section{GWs generated by seesaw phase transition}
%%%%%%%%%%%%%%%%%%%%%%%

\subsection{$B-L$ seesaw model}

%%%%%%%%%%%%%%%%%%%%%%%%%%%%%%%%%%%%%%
\begin{table}[t]
\begin{center}
\begin{tabular}{|c|ccc|c|}
\hline
      &  SU(3)$_c$  & SU(2)$_L$ & U(1)$_Y$ & U(1)$_{B-L}$  \\ 
\hline
$q^{i}_{L}$ & {\bf 3 }    &  {\bf 2}         & $ 1/6$       & $1/3 $   \\
$u^{i}_{R}$ & {\bf 3 }    &  {\bf 1}         & $ 2/3$       & $1/3 $   \\
$d^{i}_{R}$ & {\bf 3 }    &  {\bf 1}         & $-1/3$       & $1/3 $  \\
\hline
$\ell^{i}_{L}$ & {\bf 1 }    &  {\bf 2}         & $-1/2$       & $-1 $    \\
$e^{i}_{R}$    & {\bf 1 }    &  {\bf 1}         & $-1$         & $-1 $   \\
\hline
$H$            & {\bf 1 }    &  {\bf 2}         & $- 1/2$       & $0 $   \\  
\hline
$N^{i}_{R}$    & {\bf 1 }    &  {\bf 1}         &$0$                    & $-1$     \\
$\Phi_1$            & {\bf 1 }       &  {\bf 1}       &$ 0$                  & $ + 1 $  \\
$\Phi_2$            & {\bf 1 }       &  {\bf 1}       &$ 0$                  & $ + 2 $  \\ 
\hline
\end{tabular}
\end{center}
\caption{
The particle content of our $U(1)_{B-L}$ model. 
In addition to the SM particle content ($i=1,2,3$), three RH neutrinos  
  [$N_R^i$ ($i=1, 2, 3$)] and two $U(1)_{B-L}$ Higgs fields ($\Phi_1$ and $\Phi_2$) are introduced.   
}
\label{table1}
\end{table}
%%%%%%%%%%%%%%%%%%%%%%%%%%%%%%%%%%%%%%%%%%%%%%%

Our model is based on the gauge group $SU(3)_C \times SU(2)_L \times U(1)_Y \times U(1)_{B-L}$, where
 three RH neutrinos ($N_R^i$ with $i$ running $1,2,3$) and
 two SM singlet $B-L$ Higgs fields ($\Phi_1$ and $\Phi_2$) are introduced.
Under these gauge groups, three generations of RH neutrinos have to be introduced 
 for the anomaly cancellation. 
The particle content is listed in Table~\ref{table1}.
The Yukawa sector of the SM is extended to have 
\begin{align}
\mathcal{L}_{Yukawa} \supset  - \sum_{i=1}^{3} \sum_{j=1}^{3} Y^{ij}_{D} \overline{\ell^i_{L}} H N_R^j 
          -\frac{1}{2} \sum_{k=1}^{3} Y_{N^k} \Phi_2 \overline{N_R^{k~C}} N_R^k 
           + {\rm H.c.} ,
\label{Lag1} 
\end{align}
 where the first term is the neutrino Dirac Yukawa coupling, and the second is the Majorana Yukawa couplings. 
Once the $U(1)_{B-L}$ Higgs field $\Phi_2$ develops a nonzero VEV,
 the $U(1)_{B-L}$ gauge symmetry is broken and the Majorana mass terms
 of the RH neutrinos are generated. 
Then, the seesaw mechanism is automatically implemented in the model after the electroweak symmetry breaking.

We consider the following scalar potential: 
\begin{align}
V(\Phi_1, \Phi_2 )
 =& \frac{1}{2} \lambda_1 (\Phi_1 \Phi_1^{\dagger})^2+\frac{1}{2}\lambda_2 (\Phi_2\Phi_2^{\dagger} )^2
 +\lambda_3 \Phi_1\Phi_1^{\dagger} (\Phi_2 \Phi_2^{\dagger}) \nonumber \\
 & + M^2_{\Phi_1} \Phi_1\Phi_1^{\dagger} - M^2_{\Phi_2} \Phi_2\Phi_2^{\dagger} 
 - A (\Phi_1 \Phi_1 \Phi_2^{\dagger} + \Phi_1^{\dagger} \Phi_1^{\dagger} \Phi_2 ) .
\label{eq:totalpotential}
\end{align}
Here, we omit the SM Higgs field ($H$) part and its interaction terms
 for not only simplicity but also little importance in the following discussion,
 since we are interested in the case that the VEVs of $B-L$ Higgs fields are much larger than that of the SM Higgs field.\footnote{For the case 
 of a phase transition of the SM Higgs field interacting with new Higgs fields,
 see, for example, Ref.~\cite{Jinno:2015doa}.}
All parameters in the potential~(\ref{eq:totalpotential}) are taken to be real and positive.
At the U(1)$_{B-L}$ symmetry breaking vacuum, the $B-L$ Higgs fields are expanded around those VEVs $v_1$ and $v_2$, as
\begin{align}
\Phi_1 =& \frac{ v_1 + \phi_1 + i \chi_1 }{\sqrt{2}} ,\\
\Phi_2 =& \frac{ v_2 + \phi_2 + i \chi_2 }{\sqrt{2}} .
\end{align}
Here, $\phi_1$ and $\phi_2$ correspond to two real degrees of freedom as $CP$-even scalars,
  one linear combination of $\chi_1$ and $\chi_2$ is the Nambu-Goldstone mode eaten by the $U(1)_{B-L}$ gauge boson 
 ($Z'$ boson) and the other is left as a physical $CP$-odd scalar.  
Mass terms of particles are expressed as
\begin{align}
\mathcal{L}_\mathrm{mass} = & -\frac{1}{2}(\phi_2 \,\, \phi_1  )
\left(
\begin{array}{cc}
 \frac{1}{2} \lambda_3 v_1^2+\frac{3}{2} \lambda_2 v_2^2- M_{\Phi_2}^2 & v_1\left(-\sqrt{2}A+\lambda_3 v_2 \right) \\
 v_1\left(-\sqrt{2}A+\lambda_3 v_2\right) & \frac{3 \lambda_1 v_1^2}{2}+\frac{\lambda_3 v_2^2}{2}+M_{\Phi_1}^2-\sqrt{2}A v_2\\
\end{array}
\right)
\left(
\begin{array}{c}
 \phi_2 \\
 \phi_1 \\
\end{array}
\right)   \nonumber \\ & - \frac{1}{2}
(\chi_2  \,\, \chi_1  )
\left(
\begin{array}{cc}
 \frac{1}{2}\lambda_3v_1^2+\frac{1}{2}\lambda_2 v_2^2- M_{\Phi_2}^2 & -\sqrt{2}A v_1\\
 -\sqrt{2}A v_1&  \frac{1}{2}\lambda_1 v_1^2+ M_{\Phi_1}^2+ \frac{1}{2}\lambda_3 v_2^2 + \sqrt{2}A v_2 \\
\end{array}
\right)
\left(
\begin{array}{c}
 \chi_2 \\
 \chi_1 \\
\end{array}
\right)   \nonumber \\
 & - \frac{1}{2} g_{B-L}^2(4 v_2^2+v_1^2) Z^{\prime \mu} Z^\prime_\mu - \frac{1}{2}\sum_i\overline{N_i^c}\frac{Y_{N^i} v_2}{\sqrt{2}}N_i .
\end{align}
With the $U(1)_{B-L}$ symmetry breaking, the RH neutrinos $N_R^i$ and the $Z^\prime$ boson 
 acquire their masses, respectively, as 
\begin{align}
  m_{N_R^i}=& \frac{Y_{N^i}}{\sqrt{2}} v_2, \\ 
  m_{Z^\prime}^2 =& g_{B-L}^2(4v_2^2+ v_1^2),
\end{align}
where $g_{B-L}$ is the $U(1)_{B-L}$ gauge coupling. 
The mass matrix of $CP$-even Higgs bosons ($\phi_1$ and $\phi_2$) and the mass of the physical $CP$-odd scalar $P$ 
 can be, respectively, simplified as
\begin{equation}
\left(
\begin{array}{cc}
 \lambda_2 v_2^2+\frac{A v_1^2}{\sqrt{2} v_2} &  v_1 \left(\lambda_3 v_2- \sqrt{2}A \right) \\
  v_1 \left( \lambda_3 v_2-\sqrt{2}A \right) & \lambda_1v_1^2 \\
\end{array}
\right),
\label{eq:H1H2mass}
\end{equation}
and
\begin{equation}
 m_P^2 = \frac{A}{\sqrt{2}v_2}( v_1^2+4v_2^2),
\label{eq:Amass}
\end{equation}
 by eliminating $M^2_{\Phi_2}$ and $M^2_{\Phi_1}$ under the stationary conditions, 
\begin{align}
& \frac{\lambda_2 v_2^3}{2}+\frac{1}{2}\lambda_3 v_1^2 v_2 - M^2_{\Phi_2} v_2 -\frac{ A v_1^2}{\sqrt{2}} = 0,  \\
& \frac{\lambda_1 v_1^3}{2}+\frac{1}{2}\lambda_3 v_1 v_2^2 + M^2_{\Phi_1} v_1 -\sqrt{2}A v_1 v_2 = 0 .
\end{align}
Let us here note the LEP constraint $m_{Z'}/g_{B-L} =\sqrt{4 v_2^2+v_1^2} \gtrsim 6$ TeV~\cite{Carena:2004xs, Heeck:2014zfa} 
  and the constraint from the LHC Run-2  (see, for example, Refs.~\cite{Okada:2016gsh,Okada:2016tci,Okada:2017pgr,Okada:2018ktp}) 
\begin{equation}
 m_{Z'} \gtrsim  3.9~\mathrm{TeV},
\label{constr:Zpmass}
\end{equation}
 for $g_{B-L}\simeq 0.7$. 
 
\begin{figure}[htbp]
%  \begin{center}
\centering
\includegraphics[clip,width=12.0cm]{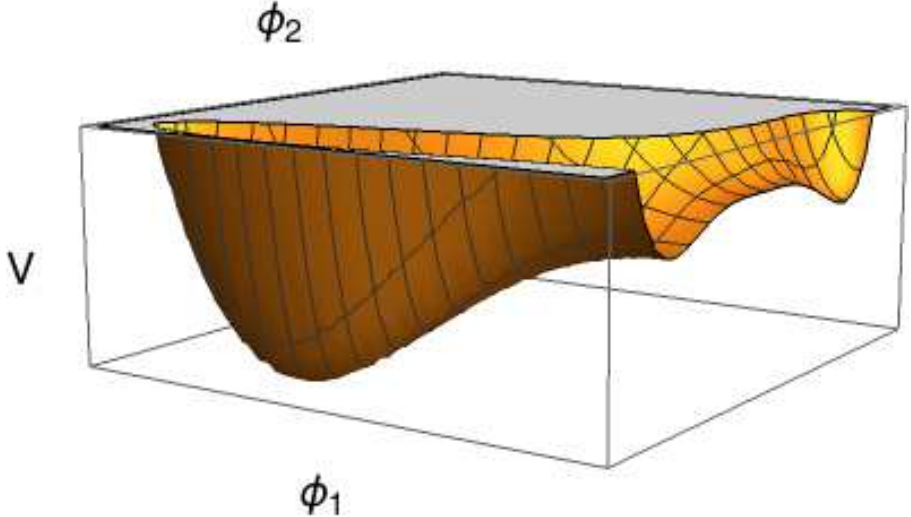}
\caption{
The three-dimensional plot of the one-loop corrected scalar potential of two $B-L$ Higgs fields
 $\Phi_1$ and $\Phi_2$ at zero temperature, which induces a first-order phase transition in the early Universe.  
}
\label{Fig:Potential}
%  \end{center}
\end{figure}

With a suitable choice of parameters, we find that 
 the phase transition of the $B-L$ gauge symmetry breaking by the Higgs fields $\Phi_1$ and $\Phi_2$
 becomes of the first order in the early Universe. 
In the following analysis, we set $\lambda_1 =0.1, \lambda_2 = 0.1$, and $\lambda_3 =0.3$
 and all corrections through neutrino Yukawa couplings $Y_{N^i}$ have been neglected,
 assuming $Y_{N^i} \ll g_{B-L}$, for simplicity.
We show in Fig.~\ref{Fig:Potential} the shape of the one-loop scalar potential (\ref{eq:totalpotential}).
%
%%%%%%%%%%%%%%%%%%%%
%%%%%%%%%%%%%%%%%%%

Implementing our model into the public code \texttt{CosmoTransitions}~\cite{Wainwright:2011kj}, 
 we have evaluated the parameters $\alpha$, $\beta$ and $T_\star$
 at a renormalization scale $Q^2=(v_2^1+v_1^2)/2$. 
We list our results for four benchmark points in Table~\ref{points}.
In Table~\ref{spectrum:mass}, we list the new particles' mass spectrum for point A, 
  which can be tested by the future LHC experiment. 
Except for point A, one can easily see the benchmark points 
 are far beyond the reach of collider experiments.   
\begin{table}
\begin{center}
  \begin{tabular}{|c|cccc|ccc|}
   \hline
 Point & $g_\mathrm{B-L}$ & $v_1$ & $v_2$ & $A$ &  $\alpha$ & $\beta/H_{\star}$ & $T_\star$   \\\hline 
A & $0.7$ & $4$ & $4$ & $1.1$ & $0.086$ & $109$ & $1.272$ \\\hline
B & $0.7$ & $100$ & $100$ & $29$ & $0.69$ & $104$ & $16.18$  \\\hline
C & $0.71$ & $10^4$ & $10^4$ & $2900$ & $0.89 $ & $52.24 $ & $1515$ \\\hline
D & $0.72$ & $10^5$ & $10^5$ & $29000$ & $0.77 $ & $57.9 $ & $15719 $ \\\hline
   \end{tabular}
\end{center}
  \caption{Input and output parameters for several benchmark points are listed. 
  All dimensionful quantities are shown in units of TeV.}
\label{points}
\end{table}
\begin{table}
\begin{center}
  \begin{tabular}{|c|cccc|}
 \hline
 Point & $m_{Z'}$ & $m_P$ & $m_{H_1}$ & $m_{H_2}$    \\\hline 
A & $6.26$ & $2.49$ & $0.65$ & $2.10$  \\\hline
%B & $15.65$ & $6.29$ & $1.44$ & $5.32$  \\\hline
   \end{tabular}
\end{center}
  \caption{The mass spectrum of the $Z'$ boson and new Higgs bosons for point A is shown in units of TeV.}
\label{spectrum:mass}
\end{table}

\begin{figure}[htbp]
%  \begin{center}
\centering
\includegraphics[clip,width=12.0cm]{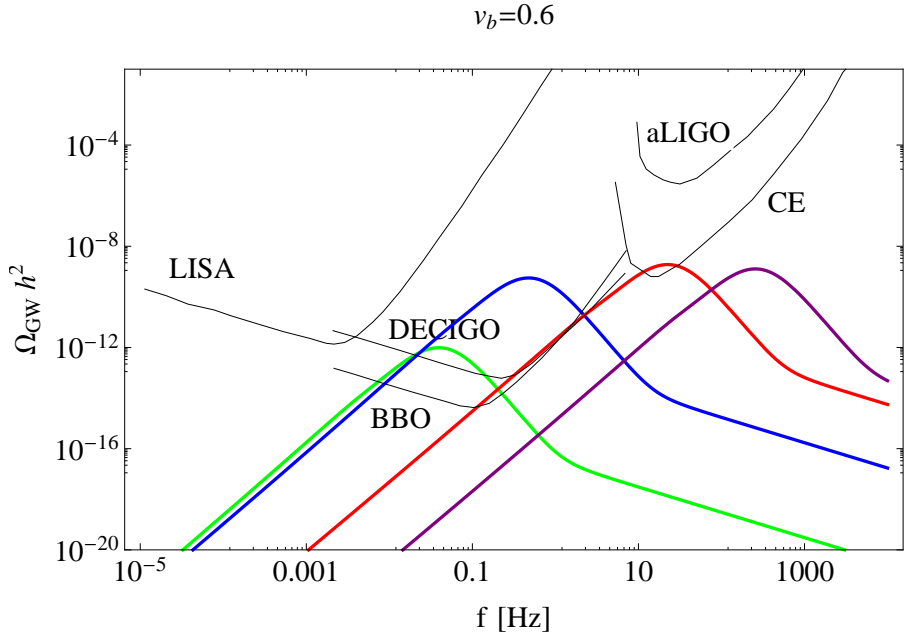}
\caption{
The predicted GW spectra for the benchmark points with $v_b=0.6$ are shown.
Green, blue, red and purple curves from left to right correspond to points A, B, C and D, respectively.
The future experimental sensitivity curves of LISA~\cite{Sathyaprakash:2009xs}, DECIGO and BBO~\cite{Yagi:2011wg},
 aLIGO~\cite{TheLIGOScientific:2014jea} and Cosmic Explore (CE)~\cite{Evans:2016mbw} are also shown 
 as black curves. 
}
\label{Fig:GWspectrum}
%  \end{center}
\end{figure}

In Fig.~\ref{Fig:GWspectrum}, we show predicted GW spectra 
 for our benchmark points along with expected sensitivities of future interferometer experiments. 
Here, the resultant spectra have been calculated with a bubble wall speed of $v_b=0.6$.
We have confirmed that the results are not so significantly changed for other $v_b$ values of $\mathcal{O}(0.1)$. 
Green, blue, red and purple curves from left to right correspond to points A, B, C and D, respectively.
Black solid curves denote the expected sensitivities of each indicated experiment,  
 according to Ref.~\cite{Sathyaprakash:2009xs} for LISA, Ref.~\cite{Yagi:2011wg} for DECIGO and BBO,
 Ref.~\cite{TheLIGOScientific:2014jea} for aLIGO and Ref.~\cite{Evans:2016mbw} for Cosmic Explore (CE).
Curves are drawn by \texttt{gwplotter}~\cite{Moore:2014lga}.
The sensitivities of DECIGO and BBO reach the results of points A and B. 
Point C is an example which is not marginally able to be detected by DECIGO/BBO 
 but its peak is within the reach of CE.  
%
%%%%%%%%%%%%%%%%%%%%
%%%%%%%%%%%%%%%%%%%

%%%%%%%%%%%%%%%%%%%%%%%
\section{Summary}
%%%%%%%%%%%%%%%%%%%%%%%

The origin of heavy Majorana RH neutrino masses is one of the essential pieces
  to understand the origin of neutrino masses through the seesaw mechanism. 
Gauged $B-L$ symmetry and its breakdown are a natural framework 
  to introduce the RH neutrinos into the SM and to generate their Majorana masses. 
The seesaw scale is in general far beyond the reach of future collider experiments. 
In this paper, we have investigated a possibility to probe the seesaw scale 
  through the observation of stochastic GW radiation.
We have shown in the context of a simple $U(1)_{B-L}$ extended SM that 
  the first-order phase transition of the $B-L$ Higgs potential 
  can generate an amplitude of GWs large enough to be detected in future experiments.
Such a detection is informative to estimate the seesaw scale. 
Grojean and Servant have shown that GWs generated by phase transitions 
  at $T_\star \sim 10^7$ GeV are in reach of future experiments~\cite{Grojean:2006bp}.(For recent studies, see e.g., Refs.~\cite{Dev:2016feu,Balazs:2016tbi}.) 
We have demonstrated that the detection of GWs is indeed possible in our model context. 

At last, we should note a delicate and critical caveat about the issue of 
  gauge dependence of the effective Higgs potential. 
See, for example, Ref.~\cite{Chiang:2017zbz} for recent discussions. 
So far, we have no clear resolution to this issue. 
According to Ref.~\cite{Chiang:2017zbz}, the resultant GW spectrum 
  has one order of magnitude uncertainties under a specific gauge choice.
Thus, even for the worst case,
 our benchmark points A and B can still be within the reach of future experiments. 
Once a better prescription has been developed, we will reevaluate the amplitude of GWs.

%======================================%
%<<<<<<<<<< ACKNOWLEDGMENTS >>>>>>>>>>>%
%======================================%

\section*{Acknowledgments}
We are grateful to C.~L.~Wainwright and T.~Matsui for kind correspondences
 concerning use of \texttt{CosmoTransitions}.
This work is supported in part by the US DOE Grant No.~DE-SC0012447 (N.O.).
%

%\newpage

%======================================%
%<<<<<<<<<<< bibliography >>>>>>>>>>>>>%
%======================================%

%%%%%%%%%%%%%%%%%%%%%%%%%%%%%%%%%%%%%%%%%%%%%%%%%%%%%%%%%%%%

%%%%%%%%%%%%%%%%%%%%%%%%%%%%%%%%%%%%%%%%%%%%%%%%%%%%%%%%%%%%

\end{document}